\documentclass[a4paper,aps,prd,10pt,preprintnumbers,twocolumn,superscriptaddress,nofootinbib,amsmath,amssymb]{revtex4-1}
\usepackage{graphicx}
\usepackage[utf8]{inputenc}
\usepackage[T1]{fontenc}
\usepackage{cmap}
\def\imo{i}

\def\K{{\cal K}}

\begin{document}
\title{Quasinormal modes of the Dirac field in the novel 4D Einstein-Gauss-Bonnet gravity}
\author{M. S. Churilova}\email{wwrttye@gmail.com}
\affiliation{Research Centre for Theoretical Physics and Astrophysics, Institute of Physics, Silesian University in Opava, CZ-746 01 Opava, Czech Republic}
\begin{abstract}
While quasinormal modes of bosonic fields for the non-trivial $4$-dimensional Einstein-Gauss-Bonnet theory of gravity suggested in [D.~Glavan and C.~Lin, Phys.\ Rev.\ Lett.\  {\bf 124}, 081301 (2020)] have been recently studied, there is no such study for fermionic fields. Here we calculate quasinormal modes of the Dirac field for spherically symmetric asymptotically flat black hole in this novel $4D$ Einstein-Gauss-Bonnet theory. The values of the quasinormal frequencies, calculated by the 6th order WKB method with Pad\'{e} approximants and the time-domain integration, show that the real part of the quasinormal modes is considerably increased, while the damping rate is usually decreasing when the coupling constant is growing.
\end{abstract}
\maketitle

\section{Introduction}

Quasinormal modes, having been observed in the modern experiments \cite{alternative1}, present the real source of information about black holes. Still these observations allow for interpretations and alternative theories of gravity \cite{alternative2}, which attempt to solve fundamental problems such as singularity problem or quantum gravitational theory. Among these alternative theories are Einstein-Gauss-Bonnet and Lovelock theories including higher curvature corrections to the Einstein term.

Recently a non-trivial Einstein-Gauss-Bonnet theory of gravity has been proposed \cite{Glavan:2019inb}, which defines a four-dimensional case as a limit~$D\!\to\!4$ of the higher dimensional one and does not require coupling to a matter field. This idea stimulated an impetuous growth of related works during last month \cite{Konoplya:2020bxa,Guo:2020zmf,Fernandes:2020rpa,Casalino:2020kbt,Wei:2020ght,Konoplya:2020qqh,Hegde:2020xlv,Kumar:2020owy,Ghosh:2020vpc,Doneva:2020ped,
Zhang:2020qew,Lu:2020iav,Konoplya:2020ibi,Singh:2020xju,Ghosh:2020syx,Konoplya:2020juj,Kobayashi:2020wqy,Kumar:2020uyz,Zhang:2020qam,Bahamonde:2020vfj,
HosseiniMansoori:2020yfj,1789072,1789085,1789073}. At the same time, while quasinormal modes of higher dimensional Einstein-Gauss-Bonnet and Lovelock theories were extensively studied \cite{Blazquez-Salcedo:2020rhf,Blazquez-Salcedo:2016enn,Konoplya:2019hml,Zinhailo:2019rwd,Konoplya:2004xx}, there is no such study for the novel $4D$ theory, except \cite{Konoplya:2020bxa}, where  quasinormal modes of scalar, electromagnetic and gravitational perturbations of spherically symmetric asymptotically flat black holes were studied. We extend these results via studying of the  Dirac field quasinormal modes of a spherically symmetric asymptotically flat black hole in this novel four-dimensional Einstein-Gauss-Bonnet theory of gravity.  This gives us description of oscillations of fermionic (such as neutrino) excitations in the vicinity of a black hole.

The eikonal instability  \cite{Dotti:2005sq,Gleiser:2005ra,Konoplya:2017lhs,Konoplya:2017zwo,Takahashi:2010ye,
Yoshida:2015vua,Takahashi:2011qda,Gonzalez:2017gwa,Konoplya:2008ix,Cuyubamba:2016cug,Takahashi:2012np} of gravitational perturbations in the novel Einstein-Gauss-Bonnet theory discussed in \cite{Konoplya:2020bxa} imposes essential constraints on the coupling constant. The obtained constraints, as it takes place in the higher dimensional Einstein-Gauss-Bonnet theory, require that the GB coupling constant is small enough \cite{Dotti:2005sq,Gleiser:2005ra,Konoplya:2017lhs,Konoplya:2017zwo,Takahashi:2010ye,Yoshida:2015vua,Takahashi:2011qda,Gonzalez:2017gwa,
Konoplya:2008ix,Takahashi:2012np}. The threshold values of the coupling constant indicate the scope for existence of a black hole, where its stability is not affected.

The paper is organized as follows. In sec. II we present the fundamentals on the Einstein-Gauss-Bonnet theory and the corresponding black hole solution. In sec. II we discuss the master wave equation. Sec. III is devoted to quasinormal modes of Dirac field. In Conclusions we summarize the obtained results and mention some open questions.

\section{The black hole metric in the novel four-dimensional Einstein-Gauss-Bonnet theory}
General Relativity in a four dimensional space-time is described by the Einstein-Hilbert action,
\begin{equation}
S_{\rm \ss EH}[g_{\mu\nu}]
	= \int \! d^{D\!}x \, \sqrt{-g}
	\left[\frac{M_{\ss \rm P}^2}{2}R\right],
\end{equation}
where $D\!=\!4$ and the reduced Planck mass~$M_{\rm \ss P}$ characterizes the gravitational
coupling strength.
The following conditions of the Lovelock's theorem \cite{Lovelock:1971yv,Lovelock:1972vz,
Lanczos:1938sf}: a) diffeomorphism invariance, b) metricity,  and c) second order equations of motion, guarantee that the Einstein theory is the unique four dimensional theory of gravity. For $D\!>\!4$ the general action corresponding to the above requirements is
\begin{equation}
S_{ \rm \ss GB}[g_{\mu\nu}]
	= \int\! d^{D\!}x \, \sqrt{-g} \, \alpha \, \mathcal{G} \, ,
\end{equation}
where~$\alpha$ is a dimensionless (Gauss-Bonnet) coupling constant and~$\mathcal{G}$ is the
Gauss-Bonnet invariant,~$\mathcal{G} \!=\!
	{R^{\mu\nu}}_{\rho\sigma} {R^{\rho\sigma}}_{\mu\nu}
	\!-\! 4 {R^\mu}_\nu {R^\nu}_\mu \!+\! R^2 \!=\!
	6 {R^{\mu\nu}}_{[\mu\nu} {R^{\rho\sigma}}_{\rho\sigma]}$.
If, as was suggested in \cite{Glavan:2019inb}, we first rescale the coupling constant
\begin{equation}
\alpha \to \alpha/(D\!-\!4) \, ,
\label{coupling}
\end{equation}
of the Gauss-Bonnet term and afterwards consider the limit~$D\!\rightarrow\!4$, we come from the solution for a static and spherically symmetric case in an arbitrary number of
dimensions~$D\!\ge\!5$ \cite{Boulware:1985wk}
\begin{equation}\label{sphansatz}
ds^2 = -f(r)dt^2
	+f^{-1}(r)dr^2
	+r^2d\Omega_{D-2}^2
\end{equation}
to the four-dimensional metric
\begin{equation}\label{sch-de}
f_\pm(r) = 1 + \frac{r^2}{\alpha }
	\Biggl[ 1\pm \biggr( \!1 \!+\! \frac{4 \alpha M}{r^3} \biggr)^{\!\! 1/2 \,} \!
	\Biggr] \, ,
\end{equation}
where $M$ is a mass parameter. Here we considered the Newton's constant $G \!=\! 1/(8\pi M_{\ss \rm P}^2)=1$ and $32 \pi \alpha$ as a new coupling constant $\alpha$. As the metric function $f_+(r)$ corresponds to asymptotically de Sitter case, here we will study $f_-(r)$, which is asymptotically flat.
Note that the black-hole metric (\ref{sch-de}) was considered earlier in \cite{Cognola,Cai:2009ua} regarding the corrections to the entropy formula.

\section{Master wave equation}

The general covariant Dirac equation has the form \cite{Brill:1957fx}:
\begin{equation}\label{covdirac}
\gamma^{\alpha} \left( \frac{\partial}{\partial x^{\alpha}} - \Gamma_{\alpha} \right) \Psi=0,
\end{equation}
where $\gamma^{\alpha}$ are noncommutative gamma matrices and $\Gamma_{\alpha}$ are spin connections in the tetrad formalism. After separation of the variables equation (\ref{covdirac}) takes the following form
\begin{equation}  \label{wave-equation}
\dfrac{d^2 \Psi_\pm}{dr_*^2}+(\omega^2-V_\pm(r))\Psi_\pm=0,
\end{equation}
where the ``tortoise coordinate'' $r_*$ is defined by the relation
\begin{equation}
dr_*= \frac{d r}{f(r)}
\end{equation}
and $\Psi_+$ is related to $\Psi_-$ by the Darboux transformation
\begin{equation}  \label{psi}
\Psi_{+}=q (W+\dfrac{d}{dr_*}) \Psi_{-}, \quad W=\sqrt{f(r)}, \quad q=const.
\end{equation}
The effective potentials can be written as follows:
\begin{equation}
V_{\pm}(r) = \frac{k}{r}f(r) \left(\frac{k}{r}\mp\frac{\sqrt{f(r)}}{r}\pm (\sqrt{f(r)})'\right),
\end{equation}
where $k=1,\;2,\;3,\;\ldots$ are multipole numbers and the prime designates the differentiation with respect to the radial coordinate $r$.

The potentials $V_+(r)$ and $V_-(r)$ can be transformed one into another by the Darboux transformation, which means that the both potentials provide the same quasinormal spectrum. Therefore we can use only $V_+(r)$ for calculation of quasinormal modes and be sure of stability of the Dirac field for the other chirality.

The effective potential $V_+(r)$ has the form of a positive definite potential barrier with a single peak.
The solutions of the master wave equation (\ref{wave-equation}) with the requirement of the purely outgoing waves at infinity and purely incoming waves at the event horizon (see,  for example, \cite{Konoplya:2011qq,Kokkotas:1999bd}) provide the discrete set of quasinormal modes.

\begin{table}
\begin{tabular}{|c|c|c|c|} \hline
  $\alpha$ & $QNM$ (WKB) &  $\alpha$ & $QNM$ (WKB) \\ \hline                                                         $-1.9$ &$0.198133 - 0.346727  i$&$-1.9$ &$0.695048 - 0.321405 i$ \\ $-1.5$ &$0.271919 - 0.304772 i$ & $-1.5$ &$0.677017 - 0.269059 i$ \\  $-1.0$ &$0.310325 - 0.259230 i$ & $-1.0$ &$0.664689 - 0.182801 i$ \\ $-0.9$ &$0.315778 - 0.252119 i$  &$-0.9$ &$0.671379 - 0.202255 i$ \\  $-0.8$ &$0.321011 - 0.245416 i$ & $-0.8$ &$0.682815 - 0.209995 i$ \\  $-0.7$ &$0.326139 - 0.239007 i$ &$-0.7$ &$0.694819 - 0.212588 i$ \\  $-0.6$ &$0.331248 - 0.232789 i$ & $-0.6$ &$0.706170 - 0.213981 i$ \\ $-0.5$ &$0.336405 - 0.226669 i$ &$-0.5$ &$0.714854 - 0.214558 i$ \\  $-0.4$ &$0.341674 - 0.220559 i$ &$-0.4$ &$0.722187 - 0.212465 i$ \\ $-0.3$ &$0.347124 - 0.214387 i$ &$-0.3$ &$0.730192 - 0.208580 i$ \\ $-0.2$ &$0.353316 - 0.207874  i$ &$-0.2$ &$0.739165 - 0.203884  i$ \\ $-0.1$ &$0.358577 - 0.200940 i$ & $-0.1$ &$0.749098 - 0.198641 i$ \\ $-0.001$ &$0.365180 - 0.193373 i$&$-0.001$ &$0.759965 - 0.192876 i$ \\ $0.001$ & $0.365338 - 0.193215 i$& $0.001$ & $0.760196 - 0.192753 i$\\ $0.05$ & $0.369254 - 0.189691 i$  &$0.05$ & $0.766022 - 0.189630 i$  \\ $0.10$ & $0.372898 - 0.186048 i$ & $0.10$ & $0.772307 - 0.186214 i$ \\ $0.15$ & $0.376604 - 0.181891 i$ &$0.15$ & $0.778970 - 0.182527 i$ \\ \hline                                                                  \end{tabular}
\caption{Dirac fundamental quasinormal mode, calculated by the WKB method, for various values of the coupling constant $\alpha$ in the stability sector; $k=1$ (left) and $k=2$ (right), $n=0$.}
\end{table}

\section{Quasinormal modes of Dirac field}

For calculating quasinormal modes we shall use the two methods in different domains:
\begin{enumerate}
\item In the frequency domain  we shall use the WKB method of Schutz and Will \cite{Schutz:1985zz}, which was extended to higher orders in \cite{Iyer:1986np,Konoplya:2003ii,Matyjasek:2017psv} and became considerably more accurate by applying Pad\'{e} approximation in \cite{Matyjasek:2017psv,Hatsuda:2019eoj}.
We use the higher-order WKB formula \cite{Konoplya:2019hlu}:
$$ \omega^2=V_0+A_2(\K^2)+A_4(\K^2)+A_6(\K^2)+\ldots- $$
\begin{equation}\nonumber
\imo \K\sqrt{-2V_2}\left(1+A_3(\K^2)+A_5(\K^2)+A_7(\K^2)\ldots\right),
\end{equation}
where $\K$ takes half-integer values. The corrections $A_k(\K^2)$ of order $k$ to the eikonal formula are polynomials of $\K^2$ with rational coefficients and depend on the values of higher derivatives of the potential $V(r)$ in its maximum. In order to increase accuracy of the WKB formula, we follow Matyjasek and Opala \cite{Matyjasek:2017psv} and use Padé approximants.

\item In the time domain we shall use integration of the wave equation at a given point in space \cite{Gundlach:1993tp} without the stationary ansatz.
We shall integrate the wave-like equation rewritten in terms of the light-cone variables $u=t-r_*$ and $v=t+r_*$. The appropriate discretization scheme was suggested in \cite{Gundlach:1993tp}:
$$
\Psi\left(N\right)=\Psi\left(W\right)+\Psi\left(E\right)-\Psi\left(S\right)-
$$
\begin{equation}\label{Discretization}
-\Delta^2\frac{V\left(W\right)\Psi\left(W\right)+V\left(E\right)\Psi\left(E\right)}{8}+{\cal O}\left(\Delta^4\right)\,,
\end{equation}
where we used the following notation for the points:
$N=\left(u+\Delta,v+\Delta\right)$, $W=\left(u+\Delta,v\right)$, $E=\left(u,v+\Delta\right)$ and $S=\left(u,v\right)$. The initial data are given on the null surfaces $u=u_0$ and $v=v_0$.
\end{enumerate}

As both methods were discussed in a numerous literature and have been recently surveyed in \cite{Konoplya:2019hlu,Konoplya:2011qq}, we shall not describe them in detail, but only demonstrate their agreement in the common range of applicability.

\begin{table}
\begin{tabular}{|c|c|}
  \hline
  $\alpha$ & $QNM$ (WKB)  \\
  \hline
  $-1.9$ &$0.376214 - 0.342049 i$ \\
  $-1.5$ &$0.994171 - 0.451980 i$ \\
  $-1.0$ &$1.025789 - 0.239301 i$ \\
  $-0.9$ &$1.037522 - 0.232209 i$ \\
  $-0.8$ &$1.048029 - 0.227364 i$ \\
  $-0.7$ &$1.058164 - 0.223339 i$ \\
  $-0.6$ &$1.068489 - 0.219562 i$ \\
  $-0.5$ &$1.079332 - 0.215782 i$ \\
  $-0.4$ &$1.090900 - 0.211856 i$ \\
  $-0.3$ &$1.103353 - 0.207650 i$ \\
  $-0.2$ &$1.116990 - 0.203065  i$ \\
  $-0.1$ &$1.131858 - 0.198150 i$ \\
  $-0.001$ &$1.148017 - 0.192669 i$ \\
  $0.001$ & $1.148360 - 0.192550 i$ \\
  $0.05$ & $1.157014 - 0.189546 i$ \\
  $0.10$ & $1.166347 - 0.186244 i$ \\
  $0.15$ & $1.176251 - 0.182660 i$ \\
  \hline
\end{tabular}
\caption{Dirac fundamental quasinormal mode, calculated by the WKB method, for various values of the coupling constant $\alpha$ in the stability sector; $k=3$, $n=0$. }
\end{table}

\begin{table}
\begin{tabular}{|c|c|c|}
  \hline
  $\alpha$ & $QNM$ (WKB) & $QNM$ (Time-domain) \\
  \hline
  \multicolumn{3}{|c|}{$k=1$}\\
  \hline
  $-1.9$ &$0.198133 - 0.346727  i$ & $0.190654 - 0.340866 i$ \\
  $-1.0$ &$0.310325 - 0.259230 i$ & $0.321342 - 0.255498 i$ \\
  $-0.5$ &$0.336405 - 0.226669 i$ & $0.347306 - 0.227229 i$ \\
  $0.15$ & $0.376604 - 0.181891 i$ & $0.376638 - 0.182514 i$\\
  \hline
   \multicolumn{3}{|c|}{$k=2$}\\
  \hline
   $-1.9$ &$0.695048 - 0.321405 i$ & $0.638486 - 0.279534 i$ \\
  $-1.0$ &$0.664689 - 0.182801 i$ & $0.682939 - 0.237913 i$ \\
  $-0.5$ &$0.714854 - 0.214558 i$ & $0.714209 - 0.216473 i$ \\
  $0.15$ & $0.778970 - 0.182527 i$ & $0.778974 - 0.182521 i$\\
  \hline
  \multicolumn{3}{|c|}{$k=3$}\\
  \hline
  $-1.9$ &$0.376214 - 0.342049 i$ & $0.977834 - 0.262061 i$ \\
  $-1.0$ &$1.025789 - 0.239301 i$ & $1.033547 - 0.231811 i$ \\
  $-0.5$ &$1.079332 - 0.215782 i$ & $1.079301 - 0.215896 i$ \\
  $0.15$ & $1.176251 - 0.182660 i$ & $1.176250 - 0.182637 i$\\
  \hline
\end{tabular}
\caption{Dirac fundamental quasinormal mode, calculated by WKB and time-domain methods, for various values of the coupling constant $\alpha$ in the stability sector; $k=1,2,3$, $n=0$.}
\end{table}

\begin{figure}[h!]
\includegraphics[width=1\linewidth]{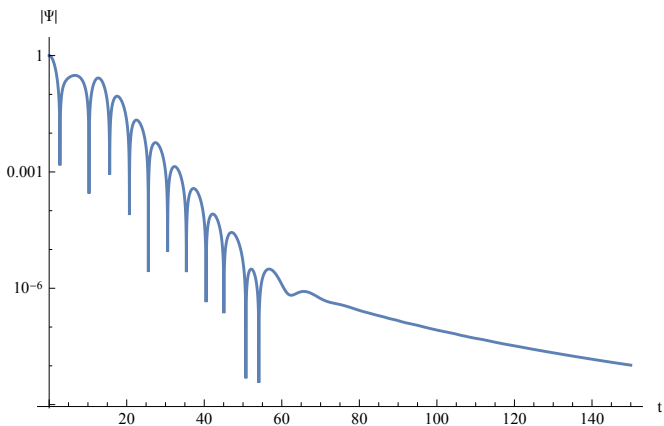}
\caption{An example of the time domain profile for the Dirac field: $k=2$, $\alpha=-1.9$.}
\label{fig1}
\end{figure}

We mainly used the 6th order WKB method with Pad\'{e} approximants \cite{Matyjasek:2017psv} for $\tilde{m} =5$ \cite{Konoplya:2019hlu} and time-domain integration to verify the obtained results. As can be seen from the Table III, the values of the quasinormal modes calculated by both methods in stability region for $\alpha$ are in a good correspondence. At $\alpha \approx -1.5$ and less there appear two concurrent modes in the time-domain profile (see an example for $k=2$ in Fig. \ref{fig1}), which makes the agreement between WKB and time-domain integration slightly worse. In this situation we are to rely on the results obtained by the time-domain integration since it is based on the convergent procedure.

From the Tables I and II one can also see that the damping rate is more sensitive to increasing the coupling constant $\alpha$ than the real oscillation frequency (which is monotonically increased). This behaviour of the quasinormal modes is qualitatively similar to the results obtained in \cite{Konoplya:2020bxa} for fields of other spin.

\begin{figure}[h!]
\includegraphics[width=1\linewidth]{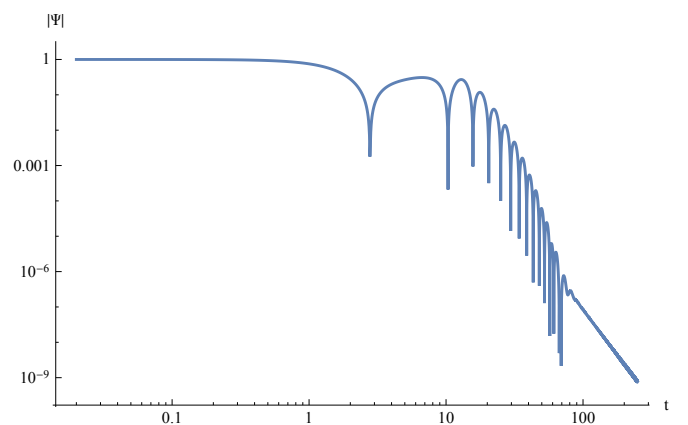}
\caption{An example of the time domain profile for the Dirac field: $k=2$, $\alpha=-0.9$.}
\label{tail}
\end{figure}

\section{Late time tails}

As the set of the quasinormal modes is incomplete, the quasinormal frequencies are suppressed by exponential or power-law tails at sufficiently late times. Fig. \ref{tail} demonstrates an example of the time domain profile for the Dirac field for the spherically symmetric asymptotically flat black hole in the novel $(3+1)$-dimensional Einstein-Gauss-Bonnet theory ($k=2$, $\alpha=-0.9$). It turned out that the late-times tails do not depend on the coupling constant $\alpha$ and for each $k$ are the same as for the Schwarzschild black hole case, satisfying the following law:
\begin{equation}
\left|\Psi\right|\sim t^{-(2 k +1)}.
\end{equation}

\section{Conclusions}

The novel formulation of the four-dimensional Einstein-Gauss-Bonnet theory is non-trivial and distinct from the pure Einstein theory even in the four dimensions owing to the re-scaling of the coupling constant. In \cite{Konoplya:2020bxa} quasinormal modes of scalar, electromagentic and gravitational perturbations of spherically symmetric asymptotically flat black hole in this theory were studied. We extended these results by calculation of the quasinormal modes of the Dirac field for the spherically symmetric asymptotically flat black hole in this novel $(3+1)$-dimensional Einstein-Gauss-Bonnet theory.

We have found that the damping rate is more sensitive to the changing of the coupling constant $\alpha$ in the stability region than the real oscillation frequency. The effect owing to the non-zero coupling constant in the stability region, especially for negative values of $\alpha$, can be considerable and reach tens of percents.

Considering of the quasinormal modes of an asymptotically de Sitter black-hole case for fields of various spin can be a further extension of our paper \cite{Churilova-progress}.

\acknowledgments{
The author acknowledges Roman Konoplya for useful discussions.}

\end{document}